\documentclass[a4paper,letters,fleqn,usenatbib,twocolumn]{mnras}

\usepackage{graphicx}	
\usepackage{amsmath}	
\usepackage{amssymb}	
\usepackage{savesym}
\usepackage{natbib}

\usepackage{mathrsfs,dsfont}
\usepackage{multirow}
\usepackage{captcont,subcaption}
\usepackage{float}
\usepackage{booktabs}
\usepackage{epstopdf}
\usepackage{color}
\usepackage{hyperref}
\usepackage{verbatim}

\usepackage[none]{hyphenat}

\graphicspath{{figures/}}

\def\be{\begin{equation}}
\def\ee{\end{equation}}

\def\Gyr{{\rm \,Gyr}}

\def\Mpc{{\rm \,Mpc}}

\def\msun{{\,M_\odot}}

\usepackage{xcolor}

\title[Collision of Merger and Accretion Shocks in Perseus]{Collision of Merger and Accretion Shocks: Formation of Mpc-scale Contact Discontinuity in the Perseus Cluster}

\author[Congyao Zhang et al.]{
Congyao Zhang,$^{1}$\thanks{E-mail: cyzhang@astro.uchicago.edu}
Eugene Churazov,$^{2,3}$
Klaus Dolag,$^{4,2}$
William R. Forman,$^5$
\newauthor
Irina Zhuravleva$^1$
\\
$^1$~Department of Astronomy and Astrophysics, University of Chicago, Chicago, IL 60637, USA \\
$^2$~Max Planck Institute for Astrophysics, Karl-Schwarzschild-Str. 1, D-85741 Garching, Germany  \\
$^3$~Space Research Institute (IKI), Profsoyuznaya 84/32, Moscow 117997, Russia \\
$^4$~University Observatory Munich, Scheinerstr 1, D-81679 Munich, Germany \\
$^5$~Smithsonian Astrophysical Observatory, Harvard-Smithsonian Center for Astrophysics, 60 Garden St., Cambridge, MA 02138 \\
}

\date{Accepted XXX. Received YYY; in original form ZZZ}

\pubyear{2019}

\begin{document}
\label{firstpage}
\pagerange{\pageref{firstpage}--\pageref{lastpage}}
\maketitle

\begin{abstract}

Two Mpc-size contact discontinuities have recently been identified in the XMM-Newton and Suzaku X-ray observations of the outskirts of the Perseus cluster (Walker et al. 2020). These structures have been tentatively interpreted as ``sloshing cold fronts'', which are customarily associated with differential motions of the cluster gas, perturbed by a merger. In this study, we consider an alternative scenario, namely, that the most prominent discontinuity, near the cluster virial radius, is the result of the collision between the accretion shock and a ``runaway'' merger shock. We also discuss the possible origin of the second discontinuity at $\sim1.2\Mpc$.

\end{abstract}

\begin{keywords}
galaxies: clusters: individual: Perseus -- galaxies: clusters: intracluster medium -- hydrodynamics -- X-rays: galaxies: clusters
\end{keywords}


\section{Introduction} \label{sec:introduction}

The Perseus cluster (A426) is the X-ray brightest cluster in the sky and it has been playing an important role in understanding the processes of active galactic nucleus (AGN) feedback \citep[e.g][]{Boehringer1993,Churazov2000,Fabian2006,Zhuravleva2014} and the properties of the hot intracluster medium \citep[ICM, e.g.][]{Hitomi2016} in its core. The outskirts of the Perseus cluster is a subject of vigorous X-ray observations
as well \citep[e.g.][]{Urban2014}. In particular, \citet{Walker2020} have recently reported the discovery of two very extended contact discontinuities (CDs) in the vicinity of the Perseus virial radius. Both of the discontinuities (at $r\sim1.2$ and $1.7\Mpc$ from NGC~1275) appear in the XMM-Newton and Suzaku data as Mpc-long tangential structures. \citet{Walker2020} suggested that these structures were generated by sloshing activity in the cluster central region some $\sim9\Gyr$ ago.

Generally, there are three most plausible scenarios leading to the formation of CDs in the context of galaxy clusters. In the first scenario, the contact discontinuity (CD) separates the low-entropy gas of the infalling subcluster from the hotter atmosphere of the main cluster \citep[e.g.][]{Vikhlinin2001}. In the second one, differential gas motions, in a stratified atmosphere of a perturbed cluster, bring gas layers with different entropies into close contact, forming a thin interface (see e.g. \citealt{Markevitch2007,Zuhone2016} for reviews), usually called a ``sloshing cold front''. The third scenario envisages a shock wave crossing another shock (or other discontinuity) and leaving behind a CD. Several versions of this last scenario have already been discussed in \citet{Birnboim2010}, \citet{Zhang2019,Zhang2020}. In this letter, we will focus on this third scenario, and argue that it can explain the main features of the observed structures in the Perseus cluster.

\section{Shock-driven CD formation scenario} \label{sec:scenario}

\begin{figure*}
\centering
\includegraphics[width=0.75\linewidth]{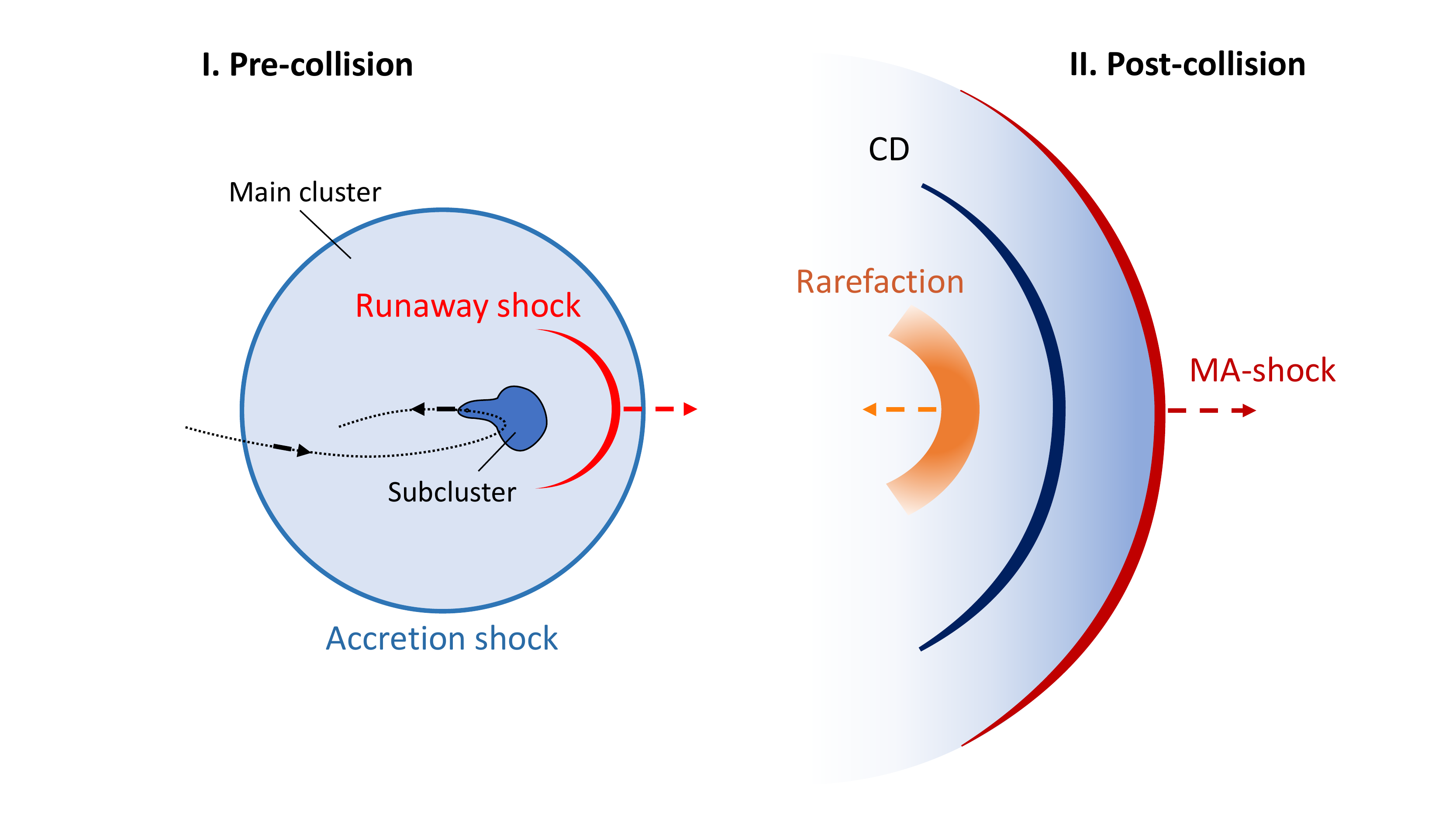}
\caption{Sketch illustrating an encounter of a runaway merger shock and an accretion shock. The black dotted line depicts the trajectory of the merging subcluster, which crosses the center of the main cluster, reaches apocenter, and then falls back. The merger shock is initially a bow shock ahead of the subcluster. At later times, the shock detaches from the subcluster and propagates ``down'' the density gradient towards the cluster outskirts. Eventually, the runaway merger shock overtakes the accretion shock, and generates three discontinuities, including a rarefaction, a contact discontinuity (CD), and a forward shock (i.e. MA-shock). The dashed arrows indicate the direction of motion of these structures. The newly found structure (at $\sim1.7\Mpc$) in the Perseus cluster may correspond to the CD shown in this sketch (see Section~\ref{sec:scenario}). }
\label{fig:sketch}
\end{figure*}

The shock-driven scenario of CD formation in cluster outskirts is illustrated in Fig.~\ref{fig:sketch}. A subcluster (i.e. dark matter halo filled with gas) moves along the trajectory depicted in the sketch as the black dotted line, and drives a bow shock ahead of it.
After the halo passes the core of the main cluster and decelerates, the shock  detaches from the subcluster and evolves into a runaway shock\footnote{See fig.~1 in \citet{Zhang2019} for a numerical experiment illustrating the transformation of a bow shock into a runaway merger shock. A signature of runaway shocks would be the identification of the driving core which could be characterized by a ``slingshot'' tail (see \citealt{Sheardown2019,Lyskova2019} for examples).}. Since the gas density profiles in the cluster outskirts are usually steep  (approximately $\propto r^{-3}$ along the non-filamentary directions; see \citealt{Vikhlinin2006}, and also \citealt{Zhang2019} and references therein), the runaway shocks are able to propagate to very large radii while maintaining their strength, and eventually encounter the accretion shocks.

Once the merger shock overtakes the accretion shock, three discontinuities are formed from smaller to larger radii, namely a rarefaction, a CD, and a forward shock. The forward shock (a.k.a. Merger-accelerated Accretion shock, MA-shock) constitutes a new boundary of the cluster atmosphere. This shock has a very high Mach number (approximately equal to the product of the Mach numbers of the runaway and accretion shocks; see the discussions in \citealt{Birnboim2010,Zhang2020}), and could travel a large distance -- up to a few virial radii into the intercluster medium before it stalls \citep[see e.g.][]{Shi2020}. What remains behind is the rarefaction and the CD. We argue that the outer structure found by \citealt{Walker2020} at  $\sim1.7\Mpc$ from the Perseus cluster center (see their fig.~2) is, in fact, such a CD.

Two facts make this scenario attractive. First, both the runaway and accretion shocks are expected to be broadly tangential to the main cluster radial direction.
And, so should be the structures formed in their collision. This is consistent with the geometry of the structure seen in Perseus. Secondly, the accretion shocks are usually located not far from the cluster virial radius \citep[see e.g.][]{Lau2015,Shi2016}. The CD formed from the shock collision appears at the initial position of the accretion shock and its radial velocity, in the rest frame of the cluster, is relatively small \citep{Zhang2020}. Therefore, the resulting CD naturally fits the properties of the outer structure found in the Perseus cluster.

\section{Illustrative simulations} \label{sec:simulation}

\begin{figure}
\centering
\includegraphics[width=0.95\linewidth]{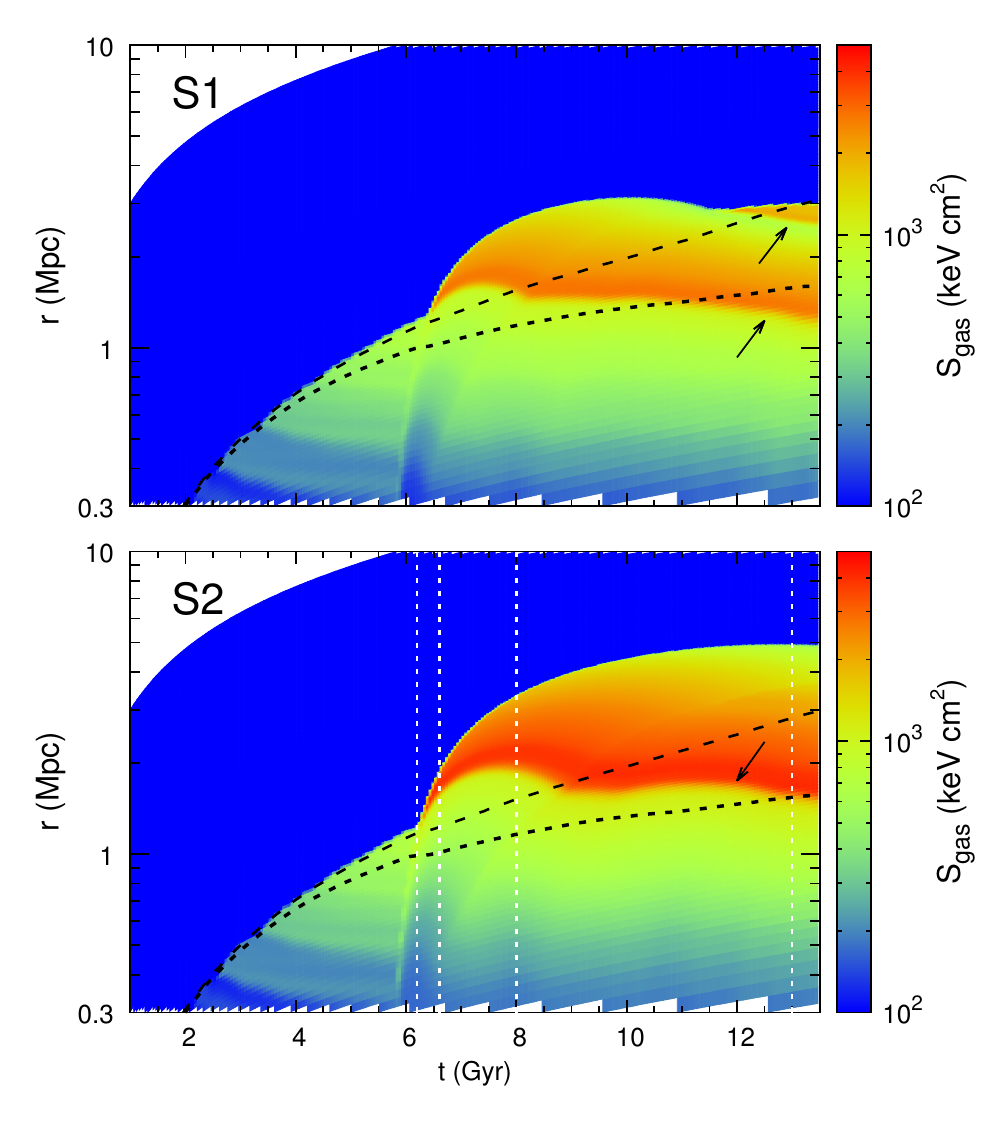}
\caption{Evolution of gas entropy profile in the 1D simulations S1 (top panel) and S2 (bottom panel). The CDs formed from the shock collisions are marked by the black arrows. The black dashed and dotted lines show the virial radii $r_{\rm 200m}$ and $r_{\rm 200c}$ of the cluster, respectively. The gas profiles at the moments marked by the white vertical lines in the run S2 are shown in Fig.~\ref{fig:profiles}. This figure illustrates the formation of CDs near the cluster virial radius, when the runaway shock overtakes the accretion shock (see Section~\ref{sec:simulation}). }
\label{fig:maps}
\end{figure}

Fig.~\ref{fig:maps} qualitatively illustrates the suggested scenario, using a spherically symmetric 1D numerical experiment. The simulation method is described in \citet{Zhang2020}, where we modelled the evolution of an idealized 1D self-similar galaxy cluster in a cosmological background. In this work, however, we applied slightly different initial conditions following that implemented in \citet{Birnboim2003}, in order to model clusters whose mass is comparable to that of Perseus\footnote{The masses of self-similar clusters used in \citet{Zhang2020} are smaller.}. We assume both the initial dark matter and gas density profiles are proportional to the two-point correlation function of the primordial cosmic density field (see appendix~C in \citealt{Birnboim2003} for more details). The amplitudes of these profiles are set so that the cluster mass is $\simeq8\times10^{14}\msun$ at the redshift $z=0$. Initially, at $z=100$, both the dark matter and gas have zero peculiar velocities. While these initial conditions are not fully self-consistent, the impact on the late evolution of a clusters is minimal.  All other parameters (e.g. spatial resolution, cosmological parameters) set in this work are the same as those used in \citet[][see their appendix]{Zhang2020}. In the simulations, an accretion shock arises naturally and its radius slowly increases with time (see Fig.~\ref{fig:maps}). At the cosmic time $t_{\rm b}=5.8\Gyr$ ($z\simeq 1$), a secondary shock is initiated at the cluster center. This shock is supposed to mimic, in the adopted 1D geometry, a runaway merger shock driven by an in-falling subcluster (see Table~\ref{tab:params} for a summary of the parameters used in the simulations). The interaction between the accretion and the ``runaway'' shocks is well resolved in these simulations.

\begin{table}
\centering
\begin{minipage}{0.45\textwidth}
\centering
\caption{Parameters of simulations (see Section~\ref{sec:simulation}).}
\label{tab:params}
\begin{tabular}{cccccc}
  \hline
  IDs & $t_{\rm b}\,(\Gyr)$\footnote{The moment a secondary shock (i.e. ``runaway'' shock) is initiated at the cluster center.}  & $\mathcal{M}_{\rm s}$\footnote{The Mach number of the runaway shock just before it collides with the accretion shock.}   \\\hline
  S1 & $5.8$ & 1.9  \\
  S2 & $5.8$ & 2.4  \\
\hline
\vspace{-6mm}
\end{tabular}
\end{minipage}
\end{table}

Fig.~\ref{fig:maps} shows the evolution of the gas entropy profile in the simulations S1 (top panel) and S2 (bottom panel). CDs formed from the shock collisions are indicated by the black arrows. Since the runaway shock develops an $N$-shaped wave profile \citep{Zhang2020}, its leading and trailing edges successively overtake the accretion shock (or MA-shock) and drive two CDs in the S1 run. As a comparison, the cluster radii $r_{\rm 200m}$ and $r_{\rm 200c}$\footnote{The $r_{\rm 200m}$ and $r_{\rm 200c}$ are the cluster radii within which the average matter density is $200$ times the mean and critical density of the Universe.} are shown as the black dashed and dotted lines in the figure, respectively. In both simulations, the CDs are formed near $r_{\rm 200m}$. The positions of MA-shocks, however, are sensitive to the Mach number of the runaway shocks. In the S2 run, the radius of MA-shock is at $\sim3r_{\rm 200c}$ at $z=0$. We note here that, while the characteristic mass scale of the simulated halo in this section is significantly larger than those shown in \citet{Zhang2020}, the simulations generally show universal features of the CD formation scenario (cf. figs.~1 and A1 in \citealt{Zhang2020}).

\begin{figure*}
\centering
\includegraphics[width=0.9\linewidth]{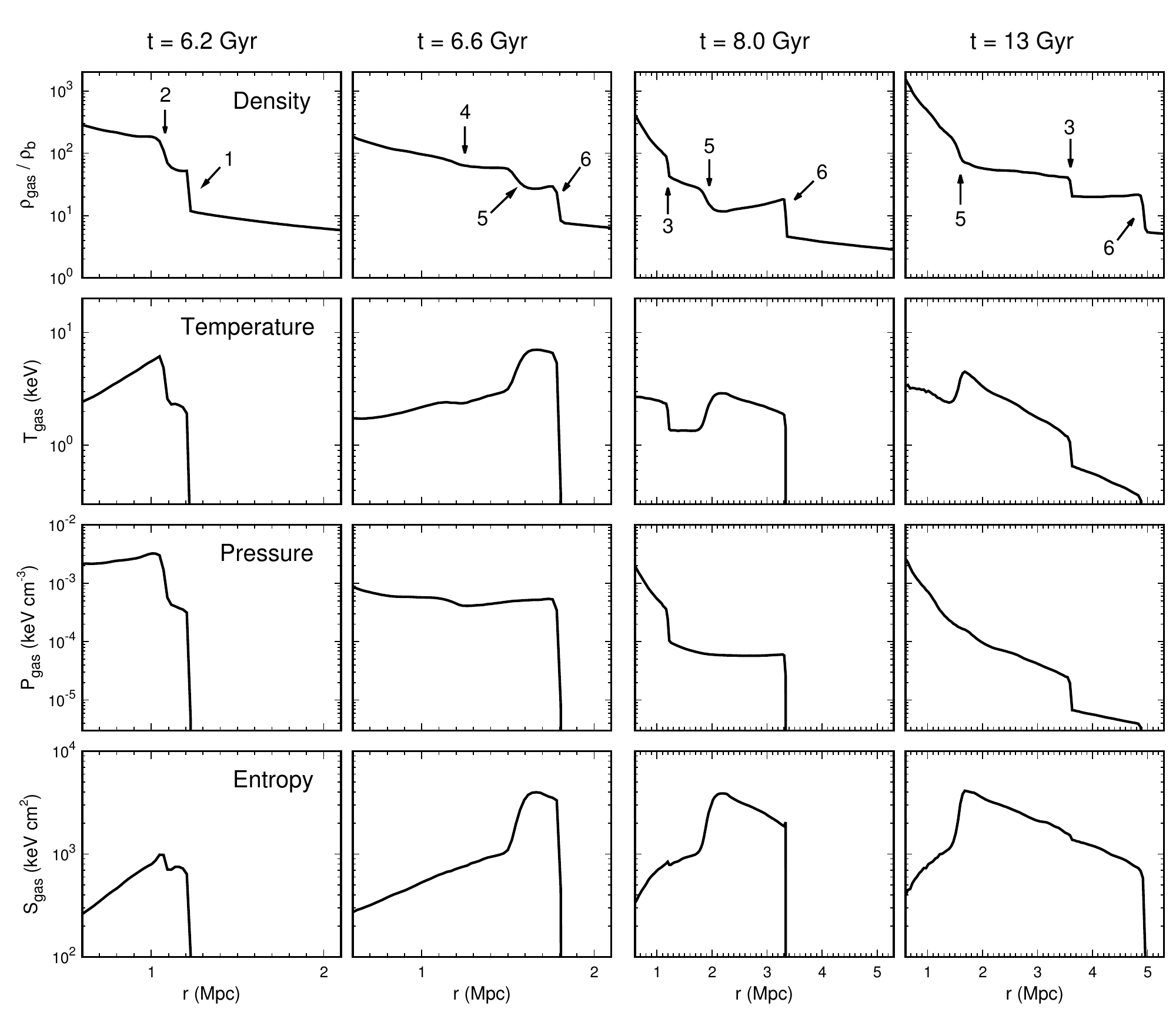}
\caption{Gas density (scaled by the cosmic mean density of baryons $\rho_{\rm b}$), temperature, pressure, and entropy profiles in 1D cosmological simulation S2 at the cosmic time $t$ indicated by the vertical white lines in Fig.~\ref{fig:maps}. The first column corresponds to the moment shortly before the runaway merger shock (its leading and trailing edges are labeled as ``2'' and ``3'', respectively) takes over the accretion shock (``1''). The remaining columns show the subsequent evolution of the profiles. After the collision, the two shocks are replaced with three new structures, namely, a rarefaction (``4''), a contact discontinuity (CD or ``cold front'', ``5''), and a forward shock (or MA-shock, ``6'').  This figure illustrates the formation scenario described in Section~\ref{sec:scenario} (cf. Fig.~\ref{fig:sketch}), and provides a good analogue to the observations (see Section~\ref{sec:simulation}).  }
\label{fig:profiles}
\end{figure*}

Fig.~\ref{fig:profiles} shows representative sets of the gas radial profiles in the simulation S2 at the cosmic time marked by the vertical white lines in Fig.~\ref{fig:maps}. The gaseous structures, including the original accretion shock, the leading and trailing edges of the runaway shock, and the ones formed from the shock collision (i.e. rarefaction, CD\footnote{Note that the smoothed appearance of the CD is mainly caused by the limited spatial resolution of the simulations (see appendix in \citealt{Zhang2020} for more information).}, and MA-shock front) are marked by the arrows in the density profiles (top panels). In particular, the first and second columns show the snapshots at the moments shortly before and after the shock collision, which provide a numerical example of the scenario depicted in Fig.~\ref{fig:sketch}. While comparing these profiles (in the three columns from right) with the observations (fig.~4 in \citealt{Walker2020}), one can see that they generally show a good match to the shape of the gas profiles around the CD. In particular, the CD separates the low- and high-entropy regions on opposite sides, whose entropies differ by a factor of $\sim 4$. This is a typical value for a CD formed by a collision of two shocks moving in the same direction where the leading shock has a much higher Mach number \citep{Birnboim2010}. \citet{Zhang2020} suggested that the high-entropy gas shell between the CD and MA-shock front is a characteristic signature of the past shocks' collision. No attempt was made to closely reproduce observed properties of the cold front in Perseus. Rather we would like to illustrate the generic scenario of the shock-driven CD formation. Detection of the MA-shock would help to confirm this scenario and constrain the relevant parameters (e.g. $t_{\rm b}$, $\mathcal{M}_{\rm s}$).

We have also revisited the full three-dimensional cosmological simulation analyzed in \citet[][see their section~3]{Zhang2020}, and found that the scenario illustrated in our 1D model generally holds in more complex situations. But to find an exact match in geometry and morphology of the structures observed in Perseus is obviously difficult in such a simulation of a single, randomly selected galaxy cluster.

Finally, we speculate on the possible origin of the inner cold front seen in Perseus ($\sim1.2\Mpc$ from the cluster center) in the scenario described above. One possibility is that it corresponds to the rarefaction formed in the same shocks' collision as the main CD (see the second column in Fig.~\ref{fig:profiles}). In observations, \citet{Walker2020} showed that the gas density and temperature jumps, for the outer cold front, are significantly larger than those for the inner one. This is broadly in line with the results seen in our simulations. Another possibility is that the observed inner structure is a fossil CD, formed by either an interaction between the accretion shock and the runaway shock driven by a past merger event, or the leading edge of the $N$-shaped runaway shock (see the top panel in Fig.~\ref{fig:maps}). \citet{Zhang2020} have shown that the trajectories of MA-shocks depend on their shock strength and the environment of galaxy clusters. It is plausible that multiple runaway merger shocks successively encounter the accretion shock (or MA-shock) in the formation history of a galaxy cluster (see fig.~7 in \citealt{Zhang2020}). Since the rarefaction corresponds to a steep pressure decrease with radius but the CD does not, accurately measuring the gas-pressure profile, e.g. via the thermal Sunyaev-Zel'dovich (SZ) effect, near the inner feature in the Perseus cluster, may help to distinguish these two possibilities.

\section{Conclusions} \label{sec:conclusion}

In this letter, we propose a shock-driven scenario for the formation of the Mpc-size CDs discovered in the outskirts of the Perseus cluster \citep{Walker2020}. In this scenario, these CDs may arise from a collision between the runaway merger shock and the accretion shock (see the sketch in Fig.~\ref{fig:sketch}).

The runaway shock is the merger shock, now detached from the subcluster that initially drives it, after apocentric passage. After the detachment, the runaway shock continues to propagate to the cluster outskirts, and eventually overtakes the accretion shock. Three discontinuities are generated in the collision between the runaway and accretion shocks, including a rarefaction, a CD, and a forward (Merger-accelerated Accretion, MA-) shock. In this scenario, the outer cold front, seen in the Perseus cluster, is explained as the CD formed in the shock collision, whose geometry and radius broadly agree with observations. The MA-shock should be located at even larger radii. Given that the merger shock velocity is obviously supersonic, the time needed for the CD formation is less than the sound crossing time of the cluster, thus alleviating the need for a long ($\sim9\Gyr$; see \citealt{Walker2020}) propagation of the cold front in the sloshing scenario, where characteristic velocities are subsonic. At the same time, the inner edge (at $\sim1.2\Mpc$) in Perseus could be explained as either the rarefaction, formed along with the outer CD, or a fossil CD, generated in a past shock-collision event. More accurate gas-pressure measurements near this structure (e.g. via the thermal SZ effect) would be helpful to understand its origin.

\section*{Acknowledgments}

WF acknowledges support from the Smithsonian Institution and the Chandra High Resolution Camera Project through NASA contract NAS8-03060. EC acknowledges support by the Russian Science Foundation grant 19-12-00369. IZ is partially supported by a Clare Boothe Luce Professorship from the Henry Luce Foundation. KD acknowledges support by the Deutsche Forschungsgemeinschaft (DFG, German Research Foundation) under Germany’s Excellence Strategy -- EXC-2094 -- 390783311.

\section*{Data Availability}

The data underlying this article will be shared on reasonable request to the corresponding author.

\bsp	
\label{lastpage}

\begin{thebibliography}{99}


\bibitem[Birnboim \& Dekel(2003)]{Birnboim2003} Birnboim, Y. \& Dekel, A.\ 2003, \mnras, 345, 349

\bibitem[Birnboim et al.(2010)]{Birnboim2010} Birnboim, Y., Keshet, U., \& Hernquist, L.\ 2010, \mnras, 408, 199

\bibitem[Boehringer et al.(1993)]{Boehringer1993} Boehringer, H., Voges, W., Fabian, A.~C., et al.\ 1993, \mnras, 264, L25

\bibitem[Churazov et al.(2000)]{Churazov2000} Churazov, E., Forman, W., Jones, C., et al.\ 2000, \aap, 356, 788

\bibitem[Fabian et al.(2006)]{Fabian2006} Fabian, A.~C., Sanders, J.~S., Taylor, G.~B., et al.\ 2006, \mnras, 366, 417

\bibitem[Hitomi Collaboration (2016)]{Hitomi2016} Hitomi Collaboration \ 2016, \nat, 535, 117

\bibitem[Lau et al.(2015)]{Lau2015} Lau, E.~T., Nagai, D., Avestruz, C., Nelson, K., \& Vikhlinin, A.\ 2015, \apj, 806, 68

\bibitem[Lyskova et al.(2019)]{Lyskova2019} Lyskova, N., Churazov, E., Zhang, C., et al.\ 2019, \mnras, 485, 2922

\bibitem[Markevitch \& Vikhlinin(2007)]{Markevitch2007} Markevitch, M., \& Vikhlinin, A.\ 2007, \physrep, 443, 1

\bibitem[Sheardown et al.(2019)]{Sheardown2019} Sheardown, A., Fish, T.~M., Roediger, E., et al.\ 2019, \apj, 874, 112

\bibitem[Shi(2016)]{Shi2016} Shi, X.\ 2016, \mnras, 461, 1804

\bibitem[Shi et al.(2020)]{Shi2020} Shi, X., Nagai, D., Aung, H., et al.\ 2020, \mnras, 495, 784

\bibitem[Urban et al.(2014)]{Urban2014} Urban, O., Simionescu, A., Werner, N., et al.\ 2014, \mnras, 437, 3939

\bibitem[Vikhlinin et al.(2001)]{Vikhlinin2001} Vikhlinin, A., Markevitch, M., \& Murray, S.~S.\ 2001, \apj, 551, 160

\bibitem[Vikhlinin et al.(2006)]{Vikhlinin2006} Vikhlinin, A., Kravtsov, A., Forman, W., et al.\ 2006, \apj, 640, 691

\bibitem[Walker et al.(2020)]{Walker2020} Walker, S.~A., Mirakhor, M.~S., ZuHone, J., et al.\ 2020, arXiv e-prints, arXiv:2006.14043

\bibitem[Zhang et al.(2019)]{Zhang2019} Zhang, C., Churazov, E., Forman, W.~R., et al.\ 2019, \mnras, 488, 5259

\bibitem[Zhang et al.(2020)]{Zhang2020} Zhang, C., Churazov, E., Dolag, K., et al.\ 2020, \mnras, 494, 4539

\bibitem[Zhuravleva et al.(2014)]{Zhuravleva2014} Zhuravleva, I., Churazov, E., Schekochihin, A.~A., et al.\ 2014, \nat, 515, 85

\bibitem[Zuhone \& Roediger(2016)]{Zuhone2016} Zuhone, J.~A., \& Roediger, E.\ 2016, Journal of Plasma Physics, 82, 535820301


\end{thebibliography}
\end{document}